\documentclass[9pt,twocolumn]{article}

\usepackage{textgreek}
\usepackage{authblk}
\usepackage{amsmath,amstext,amsfonts}
\usepackage{graphicx}
\usepackage{hyperref}
\usepackage{wrapfig}
\usepackage{multicol}

\title{Nonlinear optics in gallium phosphide cavities: simultaneous second and third harmonic generation}

\author[1]{Blaine McLaughlin}
\author[2,1]{David P. Lake}
\author[3,1]{Matthew Mitchell}
\author[1,*]{Paul E. Barclay}

\affil[1]{Department of Physics and Astronomy and Institute for Quantum Science and Technology, University of Calgary, Calgary, AB, T2N 1N4, Canada}

\affil[2]{Department of Applied Physics \& Materials Science, California Institute of Technology, Pasadena, CA, 91125, United States}

\affil[3]{Stewart Blusson Quantum Matter Institute, University of British Columbia, Vancouver, BC, V6T 1Z4, Canada}

\affil[*]{Corresponding author: pbarclay@ucalgary.ca}

\begin{document}

\maketitle




\begin{abstract}
We demonstrate the simultaneous generation of second and third harmonic signals from a telecom wavelength pump in a gallium phosphide (GaP) microdisk. Using analysis of the power scaling of both the second and third harmonic outputs and calculations of nonlinear cavity mode coupling factors, we study contributions to the third harmonic signal from direct and cascaded sum frequency generation processes. We find that despite the relatively high material absorption in gallium phosphide at the third harmonic wavelength, both of these processes can be significant, with relative magnitudes that depend closely on the detuning between the second harmonic wavelength of the cavity modes. 
\end{abstract}


\section{Introduction}

In recent years, resonant cavity structures have been used in a wide range of applications within integrated photonics. In particular, whispering gallery mode microcavities have allowed for the incorporation of nonlinear optical effects into on-chip photonics platforms \cite{li2018whispering}. The high degree of optical confinement provided by whispering gallery mode microcavities allows for integrated nonlinear devices with great efficiency and a high degree of control over the nonlinear behavior. For example, nonlinear phenomena such as second harmonic generation (SHG), sum frequency generation (SFG), third harmonic generation (THG) and Kerr frequency comb generation have been realized in lithium niobate (LN) \cite{furst2010,moore2016,luo2017,liu2017cascading,lu2019,hao2017,ye2020,he2019,gong2019,gong2020}, SiN \cite{levy2011,okawachi2011,pfeiffer2017}, AlN \cite{pernice2012,bruch2018,jung2013}, GaN \cite{roland2016,mohamed2017}, GaAs \cite{kuo2014}, GaP \cite{rivoire2009,rivoire2010,logan2018,lake2016,schneider2018,wilson2020}, and AlN/SiN composite \cite{surya2018} microcavities.

Many nonlinear photonic devices are designed for the optimization of a specific nonlinear process. However, nonlinear photonic materials often possess significant optical nonlinearities in both the second and third order. While unwanted nonlinear effects of a given order can be minimized through device design, for sufficiently large nonlinearities, both second and third-order processes contribute to a device's response. In particular, generation of the third harmonic may occur due to competing second- and third-order nonlinear processes \cite{li2018whispering}: the process of cascaded sum frequency generation (CSFG), a second-order nonlinear process that can produce a third harmonic, occurs independently of direct third harmonic generation (DTHG). These competing processes and their efficiencies will in general depend on the relative strengths of the material's nonlinear susceptibilities and the device's optical mode spectrum. A device could be designed with parameters to optimize the generation of one process over another following the theory in reference \cite{li2018whispering}. However, in a non-optimized device the contributions from each harmonic generation process may not be immediately obvious. It is therefore necessary to develop experimental methodology for discerning the contributions from each process to the third harmonic signal. 

Generation of simultaneous SH and TH signals has previously been observed in LN \cite{liu2017cascading,lin2019}, Mg:LN \cite{sasagawa2009}, and GaP \cite{schneider2018} microcavities. However, in these cases the third harmonic generation process is assumed to be either CSFG (in LN) or DTHG (in GaP). A detailed study of the generation processes of third harmonic signals in these devices has not yet been performed. Here we demonstrate the simultaneous observation of SH and TH signals generated from a telecom pump in a GaP microdisk cavity and measure their dependence on pump power and wavelength. These harmonic signals can be attributed, respectively, to a quasi-phase matched (QPM) SHG, and both DTHG and a QPM CSFG processes. The presence of CSFG is indicated by a saturation of the SH signal at high input power. Through use of coupled-mode theory, we predict that the third harmonic output power from CSFG and DTHG will scale differently with the coupled first harmonic input power. Additional analysis was performed with finite-difference time-domain (FDTD) simulations. Using the simulated whispering gallery modes (WGMs) of our microdisk devices, we calculate the nonlinear mode coupling factors that dictate the strength of the nonlinear mode interactions, and find that they are consistent with experimentally observed behaviour.

\begin{figure}[t]
\centering
\fbox{\includegraphics[width=\linewidth]{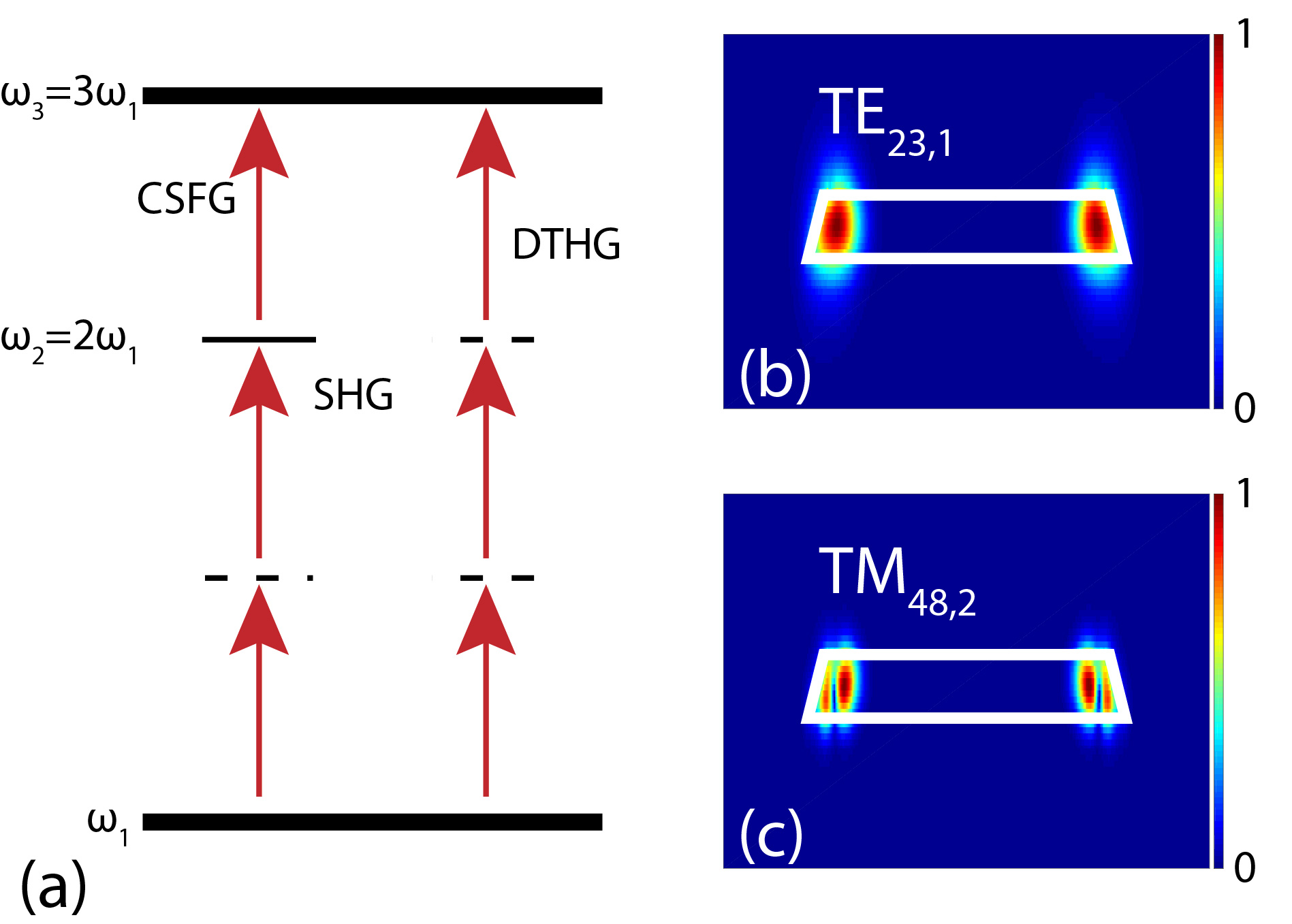}}
\caption{(a) Energy level diagram of the SHG, CSFG and THG processes. Solid lines indicate WGM resonant modes, while dashed lines represent virtual energy levels. (b) Image of FDTD simulated field magnitude of the TE\textsubscript{23,1} microdisk mode. (c) Image of FDTD simulated normalized absolute field magnitude of the TM\textsubscript{48,2} microdisk mode.}
\label{fig1}
\end{figure}

We begin in Section \ref{sec:theory} by theoretically analysing the nonlinear harmonic generation processes in gallium phosphide microdisks. In Section \ref{sec:experiment} we use this theory to analyse experimentally observed SH and TH signals from a gallium phosphide microdisk. In Section \ref{sec:discussion} we compare these results with theory, and discuss future improvements for enhancing SH and TH processes in these devices.  

\section{Theory of harmonic generation in microdisks}\label{sec:theory}

The GaP microdisk devices used in this experiment were fabricated following the process in Ref.\ \cite{mitchell2014}. The device under study has a radius of 3.05 \textmu m and a thickness of 250 nm, with a sidewall angle of 70$^{\circ}$ resulting from  the fabrication process. Previous experiments using these devices include observation of optomechanical coupling \cite{mitchell2014} and SHG from telecommunication to near-IR wavelengths \cite{lake2016}. During the SHG experiments, an additional signal corresponding to a third harmonic was also observed, but not studied in detail. A similar TH signal was observed  but not studied in detail in an independent experiment with GaP photonic crystals by Schneider et al.\ \cite{schneider2018}. The presence of simultaneous SH and TH generation indicates that despite higher loss due to material absorption in GaP at the green TH wavelength, wavelength conversion with sufficiently high efficiency is still possible. However, we can not naively assume that one of CSFG or DTHG is the process primarily responsible. 

Figure \ref{fig1}a shows an example energy diagram of the possible harmonic generation processes. Light at frequency $\omega_0$ is coupled from a waveguide into a microdisk supporting a mode with frequency $\omega_1$ close to $\omega_0$. Efficient nonlinear wave mixing processes can occur when the cavity also supports higher frequency modes whose frequencies are near the harmonics of the fundamental mode at $\omega_1$. We consider a device with modes at frequencies $\omega_2 \approx 2\omega_1$ and $\omega_3 \approx 3\omega_1$. While the excitation of photons from $\omega_1$ to $\omega_2$ will dominantly occur through SHG, excitation to $\omega_3$ can occur through two processes. The direct third-order process, DTHG, occurs independently of the SHG process assuming that we are operating with an undepleted pump. Conversely, excitation to $\omega_3$ through CSFG is an entirely second-order nonlinear process, where light from the first and SH modes undergoes sum frequency generation (SFG) to excite the TH mode. In a material with a weak third-order nonlinear susceptibility, observed TH signals can be predominantly from CSFG \cite{lin2019}. In a material with both strong second- and third-order susceptibilities, both processes can contribute significantly.

In a microcavity fabricated from a crystalline material, harmonic generation processes between modes are restricted due to polarization and phase matching requirements. The electric field of a TE(TM)-like WGM mode is given by $\mathbf{E}_{\text{TE(TM)}_{m,n}} (r,\varphi,z) \sim \mathbf{e}_{\text{TE(TM)}_{n,m}} (r,z) \exp{(im \varphi)}$, where $n$ and $m$ are the radial and azimuthal mode orders respectively. $\mathbf{e}_{TE(TM)}$ is the two-dimensional cross section of the field in the $rz$-plane. The free spectral range of microdisk devices here vary with $n$, as previously demonstrated in previous devices \cite{mitchell2014}. The fundamental TE mode in the 1550 nm range has a free spectral range of 40 nm. Higher order modes have larger free spectral range. We assume that the WGM modes in our structure with higher-order vertical index can be ignored, which is valid whenever the device is sufficiently thin \cite{borselli2006}. The modes of a microdisk differ from the perfectly polarized TE or TM modes of a cylindrical cavity \cite{jackson2001}; rather they contain non-zero field components for all dimensions, with the in-plane (i.e.\ $\hat{r}$) component resembling a typical TE field, and the out-of-plane component (i.e.\ $\hat{z}$) resembling a typical TM field. We use the polarization notation of Borselli \cite{borselli2006}, where a mode whose largest components are in-plane are considered 'TE-like', and modes whose largest components are out-of-plane are 'TM-like'. Figures \ref{fig1}b and \ref{fig1}c respectively show the simulated cross-sectional profile of the absolute value $|\mathbf{E}|$ of the TE\textsubscript{23,1} and TM\textsubscript{48,2} microdisk modes for the device studied here.

In the case of a cavity with cylindrical symmetry, the phase matching conditions of a nonlinear process can be expressed in terms of the azimuthal indices $m$ \cite{kuo2011}. Additional angular momentum quanta are accumulated due to the Berry phase of the WGM orbiting the cavity \cite{akira1986}, whose amount depends on the symmetry of the cavity material's crystal lattice \cite{lorenzoruiz2020}. GaP possesses $\overline{4}3m$ symmetry, with the crystal axis inverting with a $\pi/2$ rotation. The second-order susceptibility components will change sign under this transformation, while the third-order susceptibility components remain unchanged. As such, there is an additional factor of $\Delta m=\pm2$ added to all second-order phase matching requirements, while the third-order phase matching requirements remain unchanged. The quasi-phase matching conditions for SHG, CSFG and DTHG in GaP are therefore:
\begin{equation}
\label{eq1}
    \mathbf{SHG:} \quad m_2 = 2m_1 \pm 2,
\end{equation}
\begin{equation}
\label{eq2}
    \mathbf{CSFG:} \quad m_3=m_1 + m_2 \pm 2,
\end{equation}
\begin{equation}
\label{eq3}
    \mathbf{DTHG:} \quad m_3=3m_1.
\end{equation}
In principle, it is possible for a given set of modes with azimuthal orders $\{m_1,m_2,m_3\}$ to satisfy all three phase matching processes at once, leading to a TH signal generated both directly through phase matched DTHG and indirectly through \={4}-quasi-phase matched CSFG. 

\subsection{Coupled Mode Theory}

For nonlinear materials, the dependence of the nonlinear polarization on the electric field leads to a self-referencing wave equation \cite{boyd2008}, in which the electric field strength of the higher harmonic frequency modes depends directly on the field strength at the input fundamental frequency mode. This dependence can be expressed as a set of steady-state equations giving the field amplitudes of each optical mode in the microdisk as functions of the other optical mode amplitudes, as well as the field of the input coupling waveguide. The amplitude equation for a standing-wave mode in a waveguide coupled cavity of generic mode index $k$ is \cite{haus1984, rodriguez2007, li2018, mitchell2019, lake2020}:
\begin{equation}
    \label{eq4}
    \frac{d}{dt} a_k = \left(i\omega_k - \frac{\kappa_k}{2} \right)a_k + \sqrt{\frac{\kappa_{k,ex}}{2}} s_{+},
\end{equation}
with the waveguide transmission given by
\begin{equation}
    \label{eq5}
    s_{-} = s_{+} - \sqrt{\frac{\kappa_{k,ex}}{2}} a_k,
\end{equation} 
where $a_k$ is the electric field mode amplitude of the cavity, $s_{+(-)}$ is the incoming (outgoing) waveguide field amplitude flux, $\omega_k$ is the mode frequency and $\kappa_k$ is the mode energy decay rate. The quality factor of a mode is defined here as $Q_k \equiv \omega_k/\kappa_k$. $\kappa_k$ can be separated into contributing components
\begin{equation}
    \label{eq6}
    \kappa_k = \kappa_{k,i+p} + \kappa_{k,ex}.
\end{equation}
Here $\kappa_{k,i+p}$ is the intrinsic plus parasitic energy decay rate of the cavity, understood as the rate of energy loss into all channels other than the output mode of the waveguide, including radiation loss, optical absorption, and coupling to lossy higher-order waveguide modes \cite{spillane2003}. $\kappa_{k,ex}$ is the external energy decay rate due to the coupling between the cavity and the fundamental waveguide mode, which is used as the input and output mode. The field amplitudes are normalized such that $|a_k|^2$ is the circulating mode energy and $|s_{\pm}|^2$ is the incoming (outgoing) power flow through the waveguide. From \eqref{eq2}, the outgoing power output from the cavity mode $k$ is $P_k=(\kappa_{k,ex}/2) |a_k|^2$. 
 
To model our microdisk system, we consider a set of nearly triply resonant modes at frequencies $\omega_{1,2,3}$. The waveguide input $s_+$ is set to be nearly resonant with the fundamental mode: $\omega_0 \approx \omega_1$. Due to nonlinear effects, it can couple to the SH and TH cavity modes. Following the analyses of Rodriguez et al.\ \cite{rodriguez2007} and Li et al.\ \cite{li2018}, we obtain a set of coupled mode equations,
\begin{align}
    \label{eq7}
    \frac{d}{dt} a_1 & = \left(i\omega_1 - \frac{\kappa_1}{2} \right) a_1 + \sqrt{\frac{\kappa_{1,ex}}{2}} s_{+}, \\
    \label{eq8}
    \frac{d}{dt} a_2 & = \left(i\omega_2 - \frac{\kappa_2}{2} \right) a_2 -i \omega_2 \beta_{\text{CSFG}} a_1^* a_3 + i \omega_2 \beta_{\text{SHG}} a_1^2, \\
    \label{eq9}
    \frac{d}{dt} a_3 & = \left(i\omega_3 - \frac{\kappa_3}{2} \right) a_3 + i \omega_3 \beta_{\text{CSFG}}^* a_1 a_2 + i\omega_3 \beta_{\text{DTHG}} a_1^3,
\end{align}
where $\beta_{\text{SHG,CSFG,DTHG}}$ are the inter-modal coupling factors associated with each nonlinear frequency conversion process. For second-order nonlinear processes, the coupling factors have dimensions of $\text{J}^{-1/2}$, while for third-order nonlinear processes, the coupling factors have dimensions of $\text{J}^{-1}$. The individual terms in these equations can be understood as corresponding to photon creation for each $a_k$ term and photon annihilation for each $a_k^{*}$ term. We apply an undepleted-pump approximation, where the loss in the first harmonic mode due to the harmonic generation processes is considered negligible. This assumption is valid for low input powers on the order of \textasciitilde 1 mW \cite{li2018}.

Applying the rotating-wave approximation for each equation and solving for the steady state gives the harmonic mode output powers:
\begin{align}
    P_{2} &= \frac{\eta_{\text{SHG}} P_{in}^2}{(1+\zeta P_{in})^2},
    \label{eq10}\\
    P_{3} &= \eta_{\text{DTHG}} P_{in}^3 + \frac{\eta_{\text{CSFG}} P_{in}^3}{(1+\zeta P_{in})^2}, \label{eq11}
\end{align}
where $\eta_{\text{SHG,CSFG,DTHG}}$ is the conversion efficiency of each harmonic generation process, and $\zeta$ is a power saturation term which arises due to the CSFG process. In the special case of a triple resonance, i.e.\ $\Delta_1=\Delta_2=\Delta_3=0$ (where $\Delta_k=\omega_k-k\omega_0, \: k \in {1,2,3}$ is the frequency detuning of the resonant microdisk mode from the harmonic frequency $k\omega_0$), the efficiencies and saturation parameter are given by:
\begin{subequations}
\label{eq12}
    \begin{align}
    \eta_{\text{SHG}} & = 32 \omega_1^2 |\beta_{\text{SHG}}|^2 \frac{\kappa_{1,ex}^2\kappa_{2,ex}}{\kappa_1^4 \kappa_2^2} , \label{eq12a} \\
    \eta_{\text{DTHG}} & = 144 \omega_1^2|\beta_{\text{DTHG}}|^2 \frac{\kappa_{1,ex}^3 \kappa_{3,ex}}{\kappa_1^6 \kappa_3^2} , \label{eq12b} \\
    \eta_{\text{CSFG}} & = 2304 \omega_1^4 |\beta_{\text{SHG}}|^2 |\beta_{\text{CSFG}}|^2 \frac{\kappa_{1,ex}^3 \kappa_{3,ex}}{\kappa_1^6 \kappa_2^2 \kappa_3^2} , \label{eq12c} \\
    \zeta & = 96 \omega_1^2 |\beta_{\text{CSFG}}|^2 \frac{\kappa_{1,ex}}{\kappa_1^2 \kappa_2\kappa_3}, \label{eq12d} 
    \end{align}
\end{subequations}
where the $\beta$ terms are mode coupling factors discussed in detail below.
For low powers, the SH and TH outputs scale approximately quadratically and cubically respectively, consistent with the power scaling from pure SHG and THG. At higher input powers, the SFG process begins to deplete the SH mode energy, in a manner that scales quadratically with pump power. 

\subsection{Nonlinear Mode Coupling Factors}

Here we provide an analysis of the nonlinear coupling factors $\beta_{\text{SHG}}$, $\beta_{\text{CSFG}}$, $\beta_{\text{DTHG}}$, their dependence on the field profiles of the phase-matched modes, and their effects on the overall harmonic generation processes. The magnitudes of the coupling factors are determined by the spatial overlap of the fields of each mode participating in a given process. Following the analysis of Rodriquez et al.\ \cite{rodriguez2007}, we derive an expression for the small change in the frequency of a given microdisk mode due to nonlinear polarization. When the perturbed mode frequencies are introduced into \eqref{eq4}, we will obtain the coupled mode Eqs.\ (\ref{eq7}-\ref{eq9}) with explicit expressions for the coupling factors.

We consider a perturbation of the microdisk's dielectric constant $\delta \varepsilon$ resulting from a nonlinear optical polarization $\delta \mathbf{P}= \delta \varepsilon \mathbf{E}$. The corresponding fractional change in the mode frequency is proportional to the fraction of electric field energy in the perturbation \cite{joannopoulos2008}:
\begin{equation}
\label{eq13}
    \frac{\delta \omega_k}{\omega_k} = -\frac{1}{2} \frac{\int_{\varepsilon} \mathbf{E}^* \cdot \delta \mathbf{P}_k d^3 \mathbf{x}}{\int \varepsilon |\mathbf{E}|^2 d^3 \mathbf{x}},
\end{equation}
where $\mathbf{E}$ is the unperturbed electric field. Substituting $\omega_k \rightarrow \omega_k + \delta\omega_k$ into \eqref{eq4} for each mode $k={1,2,3}$ and rearranging for terms proportional to $a_k$ then reproduces Eqs.\ (\ref{eq7}-\ref{eq9}) with explicit expressions for each $\beta$ term given in terms of the field integrals. In order to maintain the mode amplitude normalization where $|a_k|^2$ is in units of energy, it is necessary to apply an additional normalization term $\left(\int \varepsilon |\mathbf{E}_k|^2 d^3 \mathbf{x}\right)^{1/2}$ to each $\beta$ term for each $\mathbf{E}_k$ in the numerator. This gives the nonlinear coupling factors:
%
\begin{equation}
    \label{eq14}
    \beta_{\text{SHG}} = \frac{1}{4} \frac{\int d^3 \mathbf{x} \sum_{ijk} \varepsilon \chi^{(2)}_{ijk} \left[E^{*}_{1i} E_{2j} E^{*}_{1k} + E^{*}_{1i} E^{*}_{1j} E_{2k} \right]}{\int d^3 \mathbf{x} \varepsilon |E_1|^2 \left(\int d^3 \mathbf{x} \varepsilon |E_2|^2\right)^{1/2}},
\end{equation}
\begin{equation}
    \label{eq15}
    \beta_{\text{CSFG}} = \frac{1}{4} \frac{\int d^3 \mathbf{x} \sum_{ijk} \varepsilon \chi^{(2)}_{ijk} \left[E^{*}_{2i} E_{3j} E^{*}_{1k} + E^{*}_{2i} E^{*}_{1j} E_{3k} \right]}{\left( \int d^3 \mathbf{x} \varepsilon |E_1|^2\right)^{1/2} \left(\int d^3 \mathbf{x} \varepsilon |E_2|^2\right)^{1/2} \left(\int d^3 \mathbf{x} \varepsilon |E_3|^2\right)^{1/2}},
\end{equation}
\begin{equation}
    \label{eq16}
    \beta_{\text{DTHG}} = \frac{3}{8} \frac{\int d^3 \mathbf{x} \varepsilon \chi^{(3)} \left(\mathbf{E}_1^* \cdot \mathbf{E}_1^* \right) \left(\mathbf{E}_1^* \cdot \mathbf{E}_3 \right) }{\left( \int d^3 \mathbf{x} \varepsilon |E_1|^2 \right)^{1/2} \left( \int d^3 \mathbf{x} \varepsilon |E_3|^2 \right)^{3/2}}.
\end{equation}
where $\chi^{(2)}$ and $\chi^{(3)}$ are the second- and third-order nonlinear electric susceptibilities respectively, and $\{i,j,k\}$ label the $\hat{x}$, $\hat{y}$ and $\hat{z}$ components of $\mathbf{E}$. The coupling factors can be further simplified depending on the symmetries of the device, as well as the symmetries of the nonlinear susceptibility tensors.

To estimate the values of the nonlinear coupling factors in the experiment, we have calculated the coupling factors of several potential combinations of modes obtained from FDTD simulations (performed using MEEP \cite{oskooi2010}) of WGMs in our microdisk.  The cylindrical symmetry of the microdisk allows us to specify the azimuthal mode number $m$ in the simulations. We identify the TE\textsubscript{23,1} mode to be closest fundamental mode to the target pump wavelength of 1557 nm (190 THz) used in the experiment presented below. Figure\ \ref{fig1}b shows the normalized absolute field profile of this mode. For crystals such as GaP with $\overline{4}3m$ symmetry, a TE pump field undergoing harmonic generation processes must produce a TM-like SH mode, and a TE-like TH mode \cite{boyd2008}. Additionally, SFG between a TE and a TM mode must produce a TE mode. Several TM and TE modes near the target SH and TH frequencies were simulated for phase matched azimuthal mode numbers and several radial mode numbers. Only the lowest vertical order TE and TM microdisk modes were considered in our calculations. Our simulations use a value of $n=3.01$ for all modes. Since the refractive index of GaP is dispersive in the visible regime \cite{parsons1971far}, there is considerable uncertainty in the simulated higher harmonic mode frequencies. The resulting field profiles were substituted into Eqs.\ (\ref{eq14}), (\ref{eq15}) and (\ref{eq16}) to calculate the coupling factors. Note that in general, energy matching ($\omega_2 = 2\omega_1$ etc.) is not always satisfied for a given combination of modes due to dispersion of the microdisk and GaP material. However, these calculations shed insight into relative strengths of the various coupling processes. Energy conservation is taken into account below when identifying candidate mode combinations.

\begin{figure}[ht!]
    \centering
    \fbox{\includegraphics[width=\linewidth]{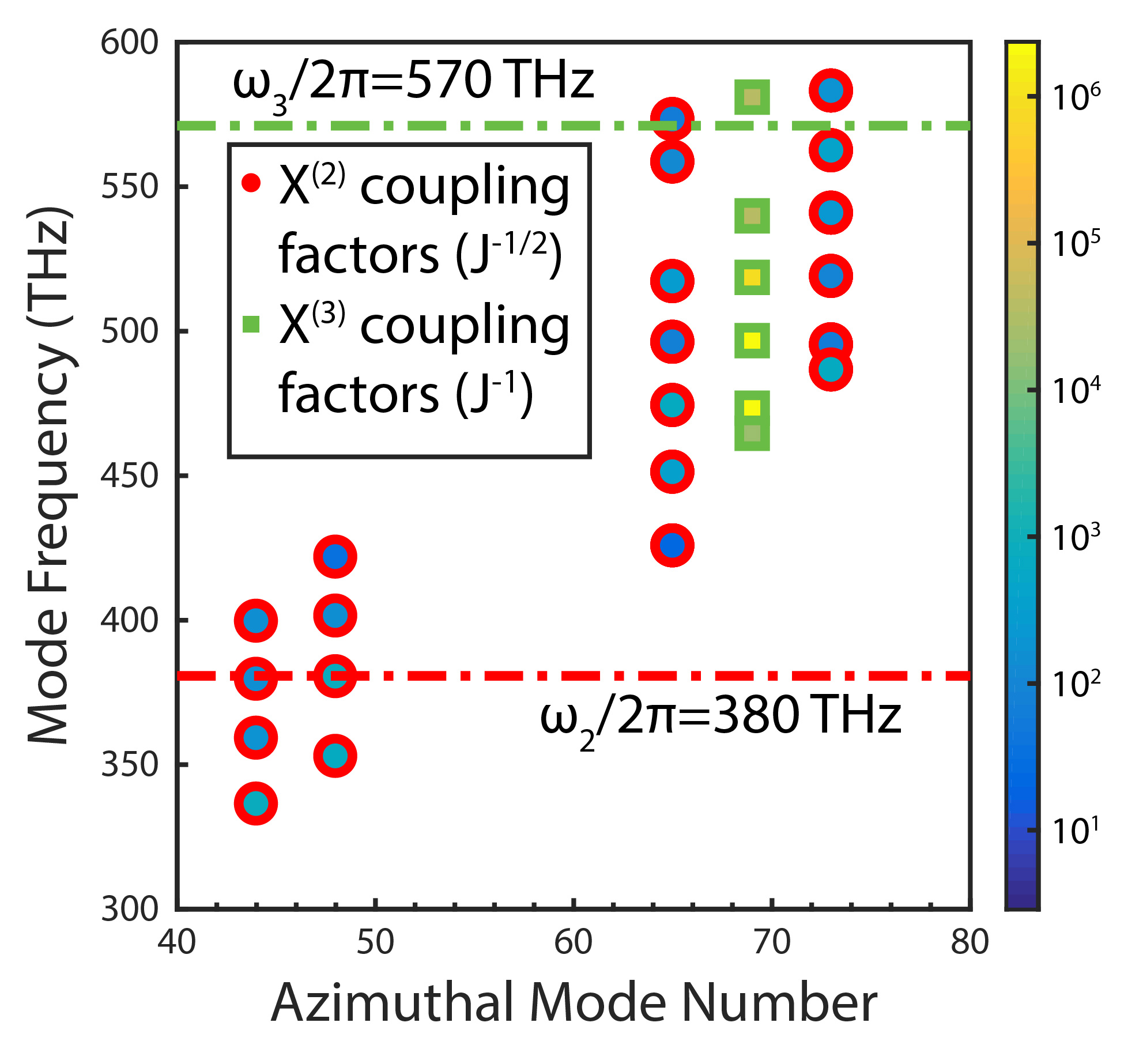}}
    \caption{Absolute values of second-order (outlined in red circles) and third-order (outlined in green squares) nonlinear coupling factors with the TE\textsubscript{23,1} first harmonic mode. Groups of phase matched microdisk modes are plotted by the frequency and azimuthal number of the highest harmonic mode in the wave mixing process.}
    \label{fig2}
\end{figure}

Figure \ref{fig2} shows the resulting calculated coupling factors for the TE\textsubscript{23,1} first harmonic mode. The mode combinations are grouped into coupling factors corresponding to second-order (highlighted in red) and third-order (highlighted in green) conversion processes. $\beta$ values are plotted in units of $\text{J}^{-1/2}$ for second-order processes and $\text{J}^{-1}$ for third-order processes, following the conventions of previous works on second and third-order harmonic generation processes in microresonators \cite{rodriguez2007,logan2018}. We obtain coupling factors on the order of $1 \sim 100 \: \text{J}^{-1/2}$ for $\beta_{SHG,CSFG}$ and $10^4 \sim 10^6 \: \text{J}^{-1}$ for $\beta_{\text{DTHG}}$. Dashed lines corresponding to the SH and TH frequencies (around 380 and 570 THz) are provided to highlight which mode combinations conserve energy. Efficient harmonic generation is predicted to occur for phase matched microdisk modes with small detunings from the harmonic frequencies.

Although it can be difficult to precisely identify which mode coupling processes are responsible for observed SH and TH signals, the magnitudes of the nonlinear coupling factors provide useful information on the relative efficiencies of the DTHG and CSFG processes. This is most useful at low input powers, where the contributions from DTHG and CSFG processes cannot be determined from power scaling. One can define a critical SH quality factor $Q_{2,c}=\frac{1}{2}|\beta_{\text{DTHG}}|/(|\beta_{\text{SHG}}||\beta_{\text{CSFG}}|)$ following the method in Ref.\ \cite{li2018}, which dictates the condition for which both processes are equally efficient. This is obtained from setting the ratio of Eqs.\ (\ref{eq12b}) and (\ref{eq12c}) to unity. DTHG typically dominates the THG process for mode combinations with $Q_2 \ll Q_{2,c}$, and likewise CSFG dominates for $Q_2 \gg Q_{2,c}$. From our coupling factor calculations, values of $Q_{2,c}$ for mode combinations that are phase matched for both DTHG and CSFG vary between $1 \sim 10^3$. Considering that previous measurements of our devices show $Q \sim 10^4$ close to the SH wavelength of 778 nm \cite{lake2016}, these results suggest that CSFG is more likely to dominate in our devices in the ideal scenario of zero detuning. However, for non-zero $\Delta_2$, $Q_{2,c}$ is reduced following the Lorentzian lineshape of the SH mode, and as we will find experimentally below, DTHG can become more significant. Note that $Q_{2,c}$ is only well defined in the case that the third harmonic mode satisfies the phase matching conditions for both CSFG and DTHG. In the case where the third harmonic mode is phase matched to only one of DTHG or CSFG, we expect that only the phase matched process may occur.

\section{Experiment}\label{sec:experiment}

To observe SH and TH signals from GaP microdisks, we used a similar setup as in our previous studies of SHG in these devices \cite{lake2016}. A fiber taper waveguide is used to evanescently couple light into microdisk modes, and to collect the SH and TH output. A continuous wave tunable laser (New Focus TLB-6700) with a central wavelength around 1557 nm was input to high-$Q$ microdisk resonances. Most high-$Q$ modes of the devices were found to exhibit SH signals, as in Ref.\ \cite{lake2016}. During these previous SHG studies, some modes were also observed to generate TH signal. Here we study a microdisk supporting a mode whose TH output was largest among measured devices, suggesting the presence of a microdisk mode close to resonance with the TH frequency. However, additional resonant spectroscopy measurements are required to confirm its precise detuning.

We perform a wavelength sweep over the pump resonance for various input powers within the range of 1-10 mW, where the saturation effects of CSFG are predicted to begin to occur \cite{li2018}. In order to minimize heating and photothermal dispersion of the microdisk mode wavelengths \cite{carmon2004}, the pump field amplitude was modulated with an electro-optic modulator (EOM) controlled by a waveform generator (Stanford DG535) outputting a square wave signal with a duty cycle of 0.5\%. The pump wavelength was independently measured using an optical spectrum analyzer (ANDO AQ6137B). The pump was then sent through an erbium-doped fiber amplifier (PriTel LNHP-FA-27-IO-CP). A variable optical attenuator (EXFO FVA-3100) was used to set the power input to the fiber taper. A set of polarization controllers were set to maximise coupling between the fiber taper field and microdisk resonances of interest. A 90:10 beamsplitter sent 10\% of the fiber taper output to a power meter (Newport 1936-R) for measuring the pump transmission, and 90\% to a spectrometer (Princeton Instruments SP2750 and Excelon PIXIS 100B CCD array). Because of the difference in power outputs at the SH and TH wavelengths, it was necessary to conduct the experiment in two sets, one for each harmonic signal. The SH signal saturated the spectrometer CCD for the input powers used here. Therefore, a neutral density filter was used at the spectrometer input to attenuate it. No such attenuation of the TH signal was required. 

\begin{figure}[ht!]
\centering
\fbox{\includegraphics[width=\linewidth]{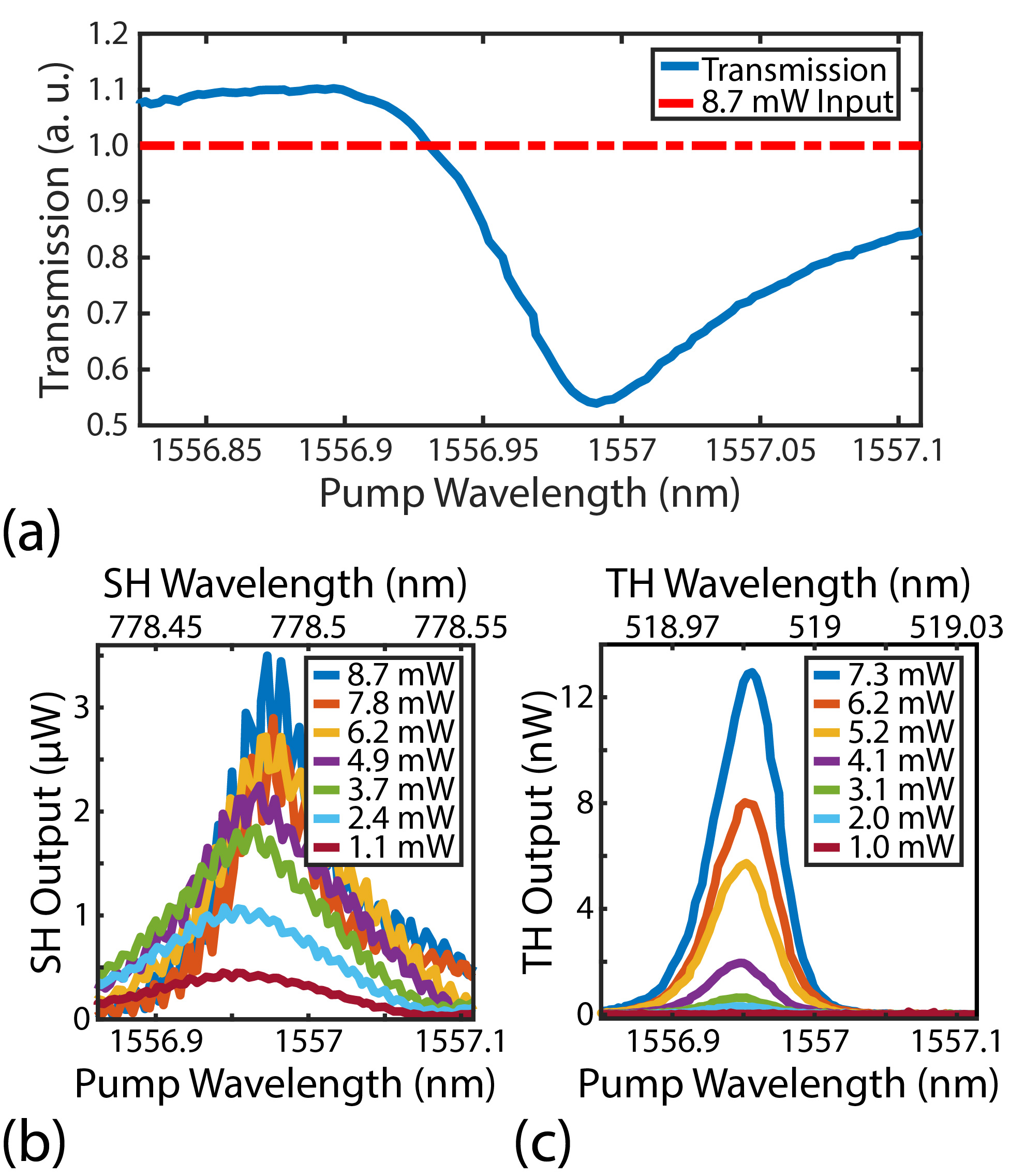}}
\caption{(a) Transmission spectrum of GaP microdisk pump mode around 1557 nm for a 8.7 mW pump power. The Fano-like lineshape of the observed resonance is attributed to a phase mismatch between the coupled fiber taper modes. (b) Second and (c) third harmonic output signal power for several pump wavelengths and input powers. Top axes show the observed peak harmonic wavelength measured in the spectrometer.}
\label{fig3}
\end{figure}

Figure \ref{fig3}a shows the transmission of the fiber taper when the pump wavelength is scanned over the first harmonic resonance for a 8.7 mW input power. A Fano lineshape was observed due the fiber taper touching the edge of the microdisk. This positioning stabilizes the fiber taper and maximizes coupling to the TH output, however, it also enhances coupling to higher-order modes of the fiber taper, resulting in a non-Lorentzian resonance lineshape \cite{wu2014}. From fits to the Fano lineshapes, we obtain the pump mode decay rates $\kappa_1 = 2.0 \times 10^{10} \: \text{rad/s}$ and $\kappa_{1,ex}=3.5 \times 10^9 \: \text{rad/s}$, corresponding to  $Q_1 \sim 6 \times 10^4$. Due to a lack of suitable tunable lasers, we were unable to directly measure the SH and TH mode spectra.

Figures \ref{fig3}b and \ref{fig3}c show the measured SH and TH output powers from the fiber taper, respectively, as a function of varying pump wavelength for several input powers. The corresponding wavelengths of the peak SH and TH signal are plotted on the top $x$-axis as a reference. From this we see that the SH and TH signals are enhanced when the pump is resonant with the cavity mode, as expected. Here the input power is defined by the power in the fiber taper immediately before the coupling region. Each point in the data set is extracted from the measured spectra of each harmonic at a given input power and finely tuned pump wavelength. A Gaussian lineshape was fit to the spectrometer resolution limited harmonic spectra, and the total number of spectrometer CCD counts under each lineshape were converted to fiber taper output power, taking into account the various efficiencies and losses of the experimental apparatus. While we attempted to limit photothermal dispersion through pulse modulation, we still observe a wavelength drift in the second harmonic spectra that increases with increasing pump power, behaving in correspondence with photothermal dispersion effects \cite{carmon2004}. The fringed dependence of the SH power on pump wavelength is related to etaloning effects in the system, which are common in back illuminated CCD detectors at near-infrared wavelengths \cite{hu2018}.

\begin{figure}[ht!]
\centering
\fbox{\includegraphics[width=\linewidth]{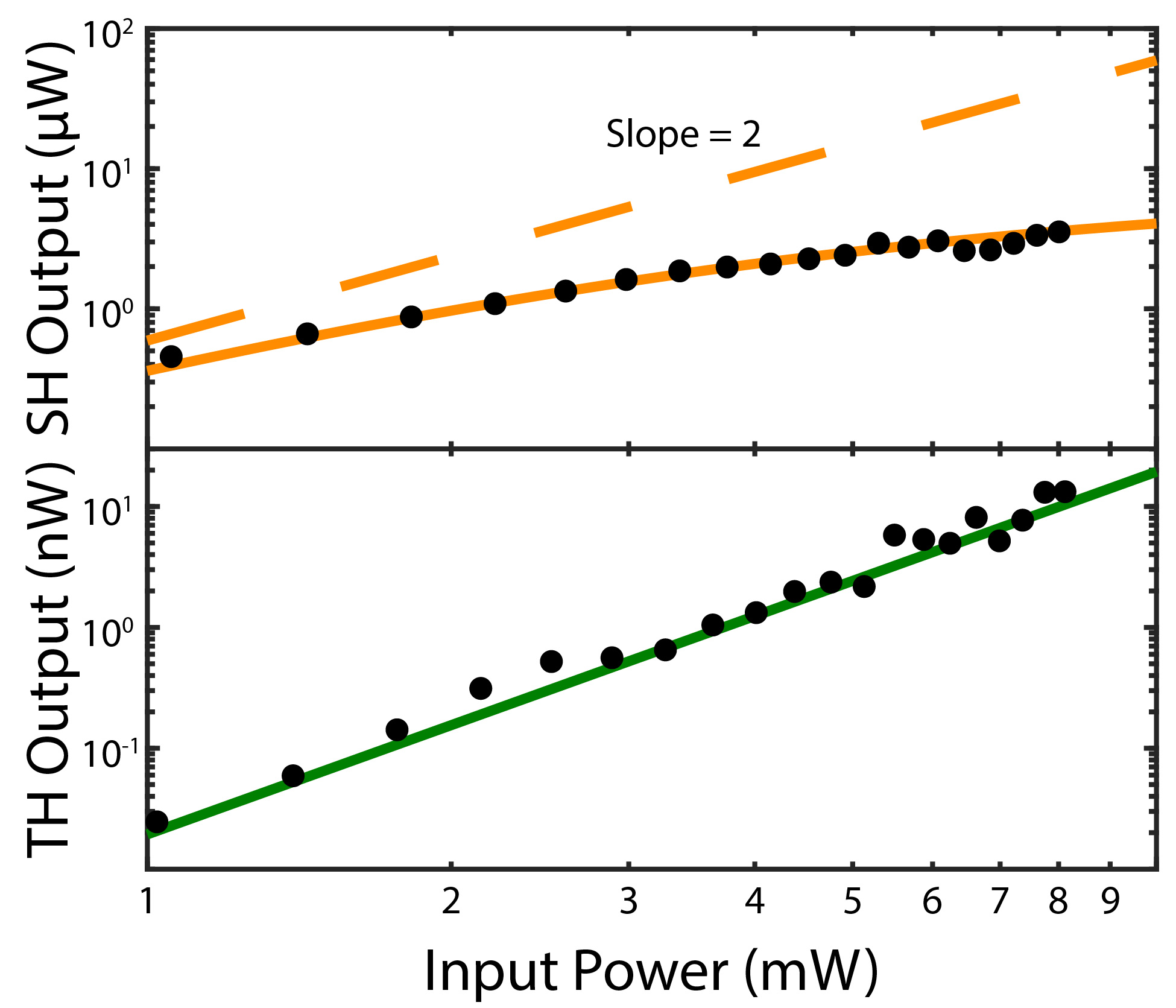}}
\caption{On-resonance output power for second (orange) and third harmonic (green) signals. Solid lines are fits to the saturated resonance theory. The dashed line corresponds to quadratic power scaling predicted from zero depletion.}
\label{fig4}
\end{figure}

Figure \ref{fig4} plots the peak SH (orange) and TH (green) output powers as a function of pump power.  The solid lines are fits to the SH and TH outputs using Eqs.\ (\ref{eq10}) and (\ref{eq11}). This data shows that the peak SH output power scales sub-quadratically with the input power, with the slope decreasing with increasing input power, indicating a saturation of the SH signal due to CSFG predicted by the denominator of \eqref{eq10}. The peak TH signal is approximately cubic with respect to the input power over the measured range, with a corresponding slope approximately equal to 3 on a log scale.

The lack of sub-cubic saturation behaviour in our measured TH signal suggests that the DTHG term in \eqref{eq11} is the main THG process in our measurements. However, this disagrees with our earlier theoretical prediction from $Q_{2,c}$ for $\Delta_2 = 0$ that CSFG should dominate. Note that pump depletion would affect both the SHG and THG signals and does not explain the observed behavior \cite{yu2017enhanced}. A plausible explanation of the difference in saturation behavior of the SHG and THG signals is that our system possesses non-zero $\Delta_2$ and small $\Delta_3$. Setting $\eta_{\text{DTHG}} = \eta_{\text{CSFG}}$ for the general case of non-zero $\Delta_2$ and $\Delta_3$ gives the condition $(Q_{2,c}/Q_2)^2((\Delta_2/\kappa_2)^2+1) = 1$. Assuming coupling to modes with $Q_{2,c} \sim 10^3$ predicted by our coupling factor calculations, then for $Q_2 \sim 10^4$, the  $\eta_{\text{DTHG}}/\eta_{\text{CSFG}} \sim 1$ condition is satisfied when $|\Delta_2| \sim 10 \kappa_2$. This value of $|\Delta_2|$ is reduced for larger $Q_{2,c}$ or smaller $Q_2$. If we assume $\Delta_3 \sim 0$, then even with $\Delta_2 \sim 10 \kappa_2$ the saturation parameter $\zeta \sim 10-20 \: \% \: \text{mW}^{-1}$ depending on the values of $|\beta_{\text{CSFG}}|$ and $\kappa_2$. As shown below, this is consistent with $\zeta$ inferred from the measured saturation of the SHG signal.

From our fits, we obtain efficiency and saturation parameters with standard errors of $\eta_{\text{SHG}} = 5.9 \: (1.6) \times 10^{-2} \:  \% \: \text{mW}^{-1}$, $\zeta = 28 \: (7) \: \% \: \text{mW}^{-1}$, and $\eta_{THG} = 1.9 \: (0.1) \times 10^{-6} \: \% \: \text{mW}^{-2}$. Our measured SHG external efficiency is of similar magnitude to what was achieved in previous SHG experiments with different microdisks on the same chip \cite{lake2016}. Our SHG and THG external efficiencies are comparable to or exceed results from similar experiments in LN and MgO:LN microdisk devices \cite{liu2017cascading,sasagawa2009}, but are smaller than current state of the art SHG-THG generation in LN microdisks \cite{lin2019}.
However, these LN devices possess $Q \sim 10^5 - 10^6$, larger than our devices. As discussed below, more efficient SHG and THG in GaP devices could be realized by reducing their optical loss, optimizing mode detunings, and improving the microdisk-waveguide output coupling efficiency.

\section{Discussion}\label{sec:discussion}

The GaP microdisks used in this experiment allow simultaneous observation of SHG and THG for relatively low powers. Our observed efficiencies and saturation are limited primarily by detuning from the SH mode, poor coupling between the SH and TH modes and the fundamental fiber taper mode, and material absorption at the TH wavelength. The fiber taper diameter is on the order of 1 $\mu\text{m}$ and provides good coupling efficiency at the pump wavelength. However, at the SHG and THG wavelengths, the coupling efficiency is reduced due to lower modal overlap of the fiber taper and microdisks fields, and the multi-mode nature of the fiber taper at these wavelengths. This suggests that the internal CSFG and DTHG conversion efficiencies are larger than what is measured in this work, resulting in a sub-quadratic scaling in the SH signal despite a low TH output power. The TH wavelength of 519 nm lies just outside the transparency window of GaP, leading to material absorption. We therefore have a fundamental lower limit on $\kappa_{3,i}$ (i.e.\ an upper limit on $Q_3$) due to absorption. The upper limit of the intrinsic quality factor of a cavity at wavelength $\lambda$ can be approximated as $Q_{i,\text{max}}(\lambda) \approx (2\pi n(\lambda))/(\lambda \alpha(\lambda))$ \cite{niehusmann2004}. Using an attenuation coefficient $\alpha_\text{GaP}(519\, \text{nm}) \approx 252 \: \text{cm}^{-1}$ \cite{borghesi1997gallium,yariv2007}, we obtain $Q_{3,i,\text{max}} \approx 1700$ or minimum decay rate $\kappa_{3,i,\text{min}} \approx 2.1 \times 10^{12} \: \text{rad/s}$. 

Using the maximum value of $Q_3$ predicted above, and assuming  $Q_2 \sim 1.0\times10^4$, as previously measured in our GaP microdisks \cite{lake2016}, for $\Delta_2 \sim \kappa_2$ and $\Delta_3 \sim 0$, we predict from Eqs.\ (\ref{eq12}) efficiency and saturation values on the same order as the measured values for coupling factors of $|\beta_{\text{SHG}}| \sim 10 \: \text{J}^{-1/2}$, $|\beta_{\text{CSFG}}| \sim 400-500 \: \text{J}^{-1/2}$ and $|\beta_{\text{DTHG}}| \sim 2-3 \times 10^6 \: \text{J}^{-1}$. These coupling factors are on the upper limit of what was calculated for $|\beta_{\text{DTHG}}|$ and $|\beta_{\text{CSFG}}|$, and on the lower limit for $|\beta_{\text{SHG}}|$.

Under ideal conditions for this device, with triple resonance and critical coupling for all harmonic modes, we estimate ideal conversion efficiencies of $\eta_{\text{SHG}} \sim 1.1 \times 10^{-1} \: \% \: \text{mW}^{-1}$, $\eta_{\text{DTHG}} \sim 3.2 \times 10^{-5} \: \% \: \text{mW}^{-2}$, $\eta_{\text{CSFG}} \sim 3.2 \times 10^{-3} \: \% \: \text{mW}^{-2}$, and $\zeta \sim 7.7 \: \% \: \text{mW}^{-1}$ for the coupling factor values listed above. These ideal values for $\eta_{\text{SHG}}$ and $\eta_{\text{CSFG}}$ are an order of magnitude larger than observed in our experiment, suggesting that the measured efficiencies are primarily limited by imperfect mode detuning, as well as undercoupling to the SH and TH modes. The maximum predicted value of $\eta_{\text{DTHG}}$ here is limited by material absorption, and may be enhanced by operating with a pump whose third harmonic wavelength lies within the transparency window of GaP. Additional resonance tuning may be achieved through the use of external temperature control \cite{koehler2018direct, logan2018} or application of external cladding \cite{thiel2021targetwavelengthtrimmed}. The ideal predicted value of $\zeta$ is lower than what was measured here, which can be attributed to undercoupling to the SH and TH modes.
 
In future experiments, both DTHG and CSFG in a GaP microcavity device could be enhanced in several ways. In addition to the techniques mentioned above, optimizations in fabrication processes \cite{honl2018highly}, reductions of sidewall roughness, improvements to waveguide coupling efficiency, dispersion engineering to ensure triple resonance and maximizing constructive DTHG-CSFG interference \cite{li2018}, and gain-induced loss compensation with dopants \cite{yu2017enhanced} would significantly improve the device performance. Alternatively, the use of photonic crystals \cite{rivoire2011second} or topological metamaterials \cite{smirnova2019third,wang2019topologically,kruk2019nonlinear} may yield improved performance due to lower mode volumes.

In summary, we have observed the simultaneous generation of SH and TH signals from a telecom pump in GaP microdisks. This work represents the first detailed experimental analysis of third harmonic generation in GaP microcavities. For experiments in which simultaneous second and third generation is observed, the methods presented here will allow systematic analysis of the third harmonic generation processes. Future third harmonic generation experiments in GaP microcavity devices should to be performed in carefully designed integrated devices whose modes are optimized for the desired wavelength and power range \cite{li2018}. 

© 2022 Optica Publishing Group. One print or electronic copy may be made for personal use only. Systematic reproduction and distribution, duplication of any material in this paper for a fee or for commercial purposes, or modifications of the content of this paper are prohibited.

\section{Funding}
This work was supported by the Natural Sciences and Engineering Research Council of Canada (Discovery Grant and Research Tools and Instruments programs). 
\section{Acknowledgments}
The authors thank D.\ Sukachev for their input.
\section{Disclosures}
The authors have no competing interests to declare.
\section{Data Availability Statement}
Data underlying the results presented in this paper are not publicly available at this time but may be obtained from the authors upon reasonable request.

\bibliographystyle{plain}
\bibliography{references}



\end{document}